\documentclass{jpsj2}
\usepackage{graphics}

\def\runtitle{Ordered Phases and Phase Transitions
in The Stacked Triangular Ising Antiferromagnet CsCoCl$_{3}$ and CsCoBr$_{3}$}
\def\runauthor{Norikazu \textsc{Todoroki} and Seiji \textsc{Miyashita}}

\title{Ordered Phases and Phase Transitions \\
in The Stacked Triangular Antiferromagnet CsCoCl$_{3}$ and CsCoBr$_{3}$}

\author{%
Norikazu \textsc{TODOROKI}\thanks{todoroki@spin.t.u-tokyo.ac.jp} and
Seiji \textsc{MIYASHITA}
}

\inst{%
Department of Applied Physics, University of Tokyo, Bunkyo-ku, Hongo 7-3-1,
Tokyo 113-8656, Japan
}

\recdate{\today}

\abst{%
We study the ordered phases and the phase transitions 
in the stacked triangular antiferromagnetic Ising (STAFI) model 
with strong interplane coupling modeling CsCoCl$_3$ and CsCoBr$_3$.
We find that there exists an intermediate phase which consists of a single 
phase of so-called partial disordered (PD) type,
and confirm the stability of this phase.
The low temperature phase of this model is so-called two-sublattice ferri magnetic phase.
The phase transition between the PD phase and two-sublattice ferri magnetic phase is of the first order.
This sequence of the phases is homomorphic as that in the three-dimensional generalized 
six-state clock model which have the same symmetry of the STAFI model.
By studying distributions of domain walls in one dimensional chains connecting layered triangular lattices,
we clarify the nature of the phase transition and give an interpretation of little anomaly of the specific heat.
}

\kword{%
CsCoCl$_3$, CsCoBr$_3$, quasi-one dimensional Ising model, frustration, partial disordered phase
}

\begin{document}
\sloppy
\maketitle

\section{Introduction}
The hexagonal ABX$_3$-type crystals CsCoCl$_{3}$ and CsCoBr$_{3}$
 have attracted a great interest because of the successive phase transitions 
due to strong frustration. 
For a long time, a large number of experimental and theoretical studies
have been done for this type of substances.
The magnetic ion Co$^{2+}$ has fictitious $S=1/2$ Ising spin.
These substances consist of 
a quasi-one-dimensional spin chains
because the exchange interaction along the c-axis is much larger than that in the 
c-plane, \cite{experiments}
 and they are coupled via antiferromagnetic interaction in the c-plane,
This type of substances is expressed by the stacked triangular 
antiferromagnetic Ising (STAFI) model.

\begin{figure}
\includegraphics[scale=0.8]{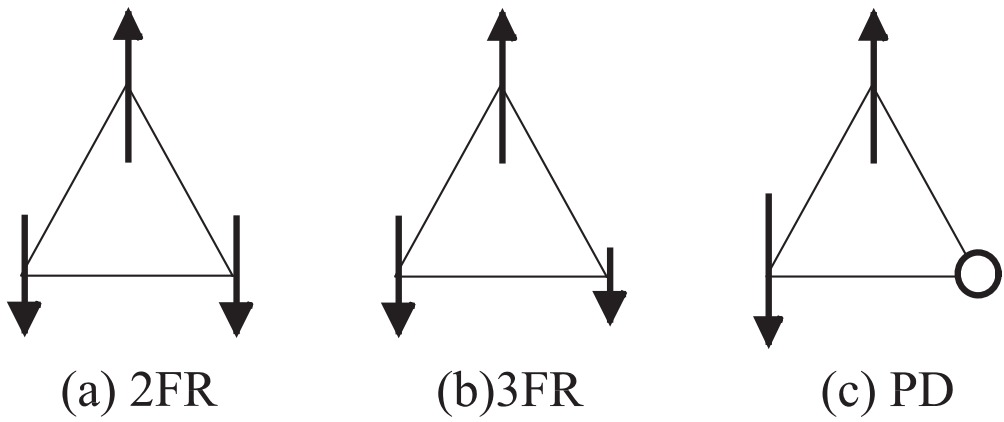}
\caption{Schematic sublattice magnetization of (a) 2FR phase (b) 3FR phase (c) PD phase.\label{submag}}
\end{figure}

The neutron scattering experiments on CsCoCl$_3$ \cite{neutron1} and CsCoBr$_3$ \cite{neutron2} suggested that
this type of substances has the two magnetic phase transitions. 
In order to explain those experimental results, the theoretical study on 
this model was done by Mekata.\cite{mekata}
Due to the strong interaction along the c-axis,
one may assume that, in the ordering process, the spins in a chain can be regarded as 
a magnetic moment, and these magnetic moments form a triangular lattice. 
Mekata studied the phase diagram of this two-dimensional model 
by using a mean field approximation (MFA).
He introduced the next-nearest neighbor interaction in the c-plane 
 to explain the successive phase transitions observed in experiments.
He found that this model has three type of ordered phases. That is,
as the temperature is reduced, there appear
the partial disordered (PD) phase, three-sublattice ferrimagnetic (3FR) phase and two-sublattice ferrimagnetic (2FR) phase 
which are schematically depicted in Fig.\ref{submag}. 
In the PD phase, the spins on one of the three sublattices are disordered and the spins
on the other two sublattices are ordered antiferromagnetically. In the 3FR phase,
the spins on two sublattices are ordered in the same direction and the spins on the last sublattice is 
ordered in the opposite direction. 
The magnitudes of the sublattice magnetizations are mutually 
different. 
In the 2FR phase, the magnitudes of the parallel sublattice magnetizations are equal. 
The temperature dependence of the magnetic structure factors are in good agreement 
with the experimental observations.
However, in the experiments, the specific heat has no anomaly at the low temperature transition point. 
As to the nature of the intermediate phase, many possibilities have been proposed,
and the nature of the intermediate phase has been mystery for a long time.

\begin{figure}
\includegraphics[scale=0.8]{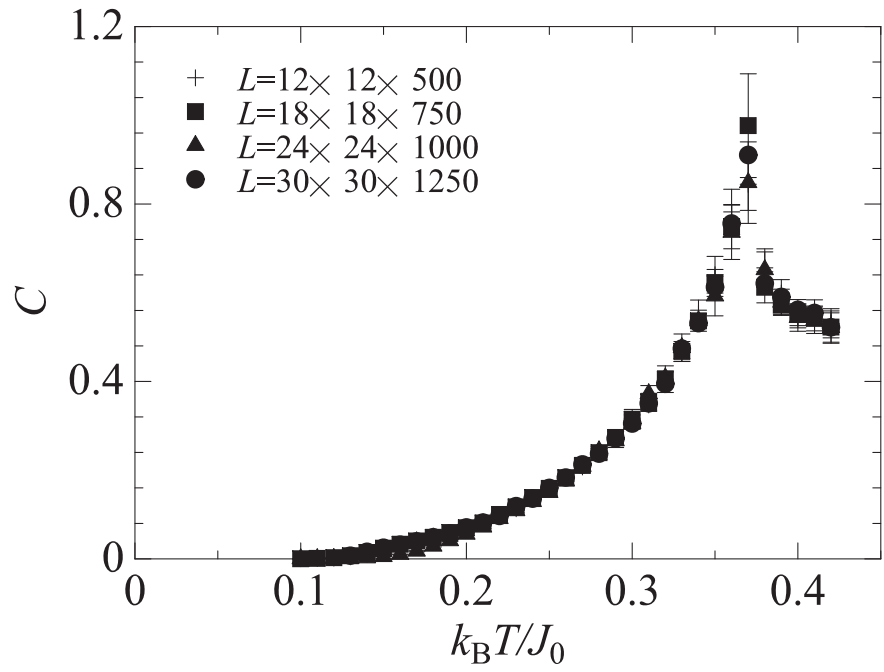}
\caption{Temperature dependence of the specific heat. \label{koseki}}
\end{figure}

In the numerical studies, 
it had been difficult to simulate the STAFI model by using a single spin flip Monte Carlo (MC) method owing to its frustration and strong one-dimensional anisotropy.
Recently, Koseki and Matsubara conquered this difficulty by introducing the cluster heat bath (CHB) 
algorithm. \cite{koseki}
They obtained the results that agree with the experiments. 
Their results reproduce experimental data about the magnetic structure factors and the specific heat (see Fig.\ref{koseki}). 
However, in their results, 
there were large fluctuations in the intermediate temperature phase corresponding to the PD phase.
Thus, the nature of the intermediate phase and property of the low temperature phase transition have been not clear.
In particular, at low temperature, the correlation length along the c-axis in the disordered chains is very long and 
it seems to diverge in simulations in small system sizes. Therefore, it could be possible that the low temperature phase transition may not be a transition in equilibrium but a crossover for the finite size effect.

From the view point of the symmetry, the STAFI model has the Z$_6$ symmetry\cite{fujiki} and the behavior of the phase transition of this model may be explained by the three dimensional six-state clock model which has also the Z$_6$ symmetry.
The simple six-state clock model has been found to have only one type of symmetry broken low temperature phase. \cite{miyashita}
On the other hand, it turned out that a generalized six-state clock (6GCL) model has 
an intermediate phase where two neighboring states are mixed (incompletely ordered phase, IOP).
In this paper, we propose that the structure of phases in the 6GCL is the same as that of the present STAFI model.
In the 6GCL, phase transition 
between the intermediate phase the paramagnetic phase belongs to the universality class of three dimensional XY model,
 and the phase transition between the IOP
and the low temperature phase is of the first order.\cite{todoroki}
Therefore, we expect, from the correspondence between the STAFI model and 
the 6GCL model, that the intermediate phase on the STAFI model is the PD phase,
and the phase transition between the PD phase and the low temperature phase is of the first order. 
It has been known that the intermediate temperature phase of the 6GCL model consists of a 
single phase of IOP, and thus we expect a single intermediate temperature phase,
i.e., the PD phase, exists in the STAFI model. That is,
the complicated phases so far proposed or the 3FR phase do not appear.
In the present paper, 
we investigate characteristics of the low temperature phase transition 
by using the MC simulation and cluster mean field approximation.

\section{Model and Numerical Results}
\begin{figure}
\includegraphics[scale=0.8]{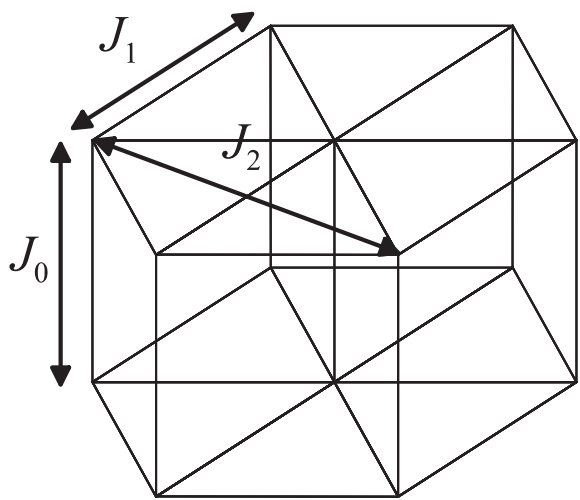}
\caption{The exchange interactions in the stacked triangular lattice.\label{sti}}
\end{figure}

The Hamiltonian of the STAFI model is given by
\begin{eqnarray}
{\cal H}=-\frac{J_0}{2}\sum_{i}S_{i,\mu}S_{i+1,\mu}
-\frac{J_1}{2}\sum_{\langle \mu,\nu\rangle }^{nn}S_{i,\mu}S_{i,\nu}
-\frac{J_2}{2}\sum_{\langle \mu,\lambda\rangle }^{nnn}S_{i,\mu}S_{i,\lambda},
\end{eqnarray}
where $S_{i,\nu}=\pm 1$. $\sum^{nn}$ and $\sum^{nnn}$ run over nearest neighbor pairs 
and next nearest neighbor pairs in the c-plane of the stacked triangular lattice (Fig.\ref{sti}), respectively. 
The sign of the exchange interactions $J_0$ and $J_1$ are negative while $J_2$ is positive.
We set exchange interaction to $|J_1/J_0|=3\times 10^{-2}$,
$|J_2/J_0|=1\times 10^{-4}$ which are the same as those used by Koseki and Matsubara.\cite{koseki}
We use lattices of the size $6n \times 6n \times 250n$ with $n=4\sim 7$. 
We impose the periodic boundary condition for all directions.
We perform MC simulation with the CHB algorithm. \cite{chb}
The initial $1000\times n$ steps are used for equilibrating process. 
After this, $5000\times n$ steps are used to take thermal averages of physical quantities.

\begin{figure}
\includegraphics[scale=0.8]{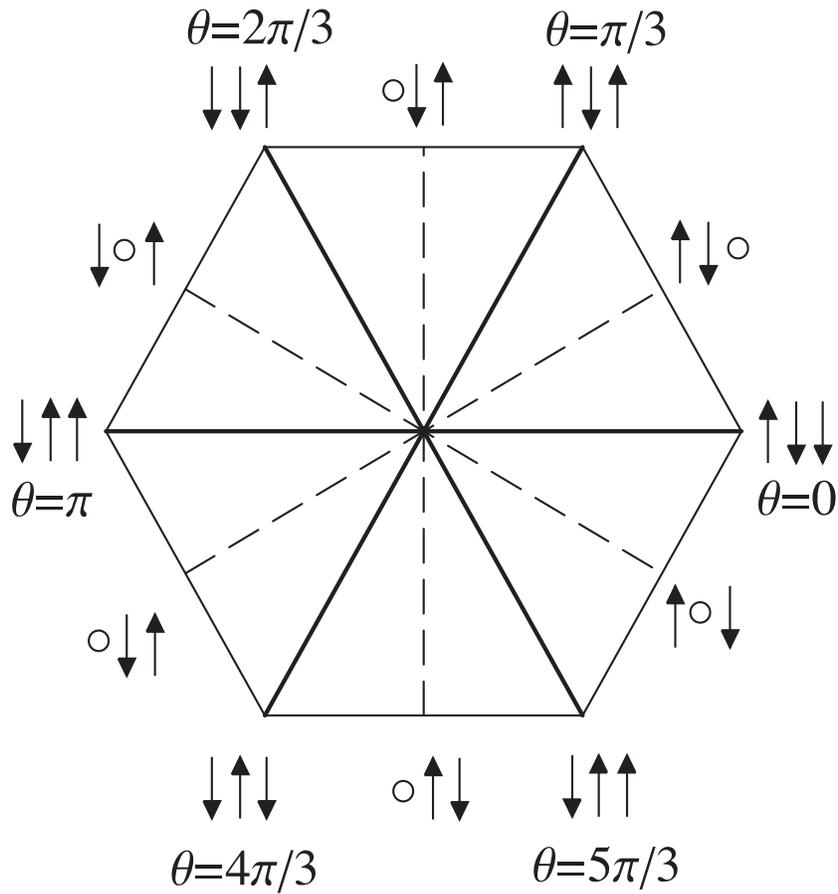}
\caption{Order parameter space of the STAFI model. The solid line connecting a vertex and the center is one of the six low temperature states
and dotted line between the neighboring solid lines is one of the six PD states. 
The three arrows denote the sublattice magnetizations: $(m_1,m_2,m_3)$.
The angles are assigned by the relation \ref{theta}\label{clock}}
\end{figure}

\begin{figure}
\includegraphics[scale=0.8]{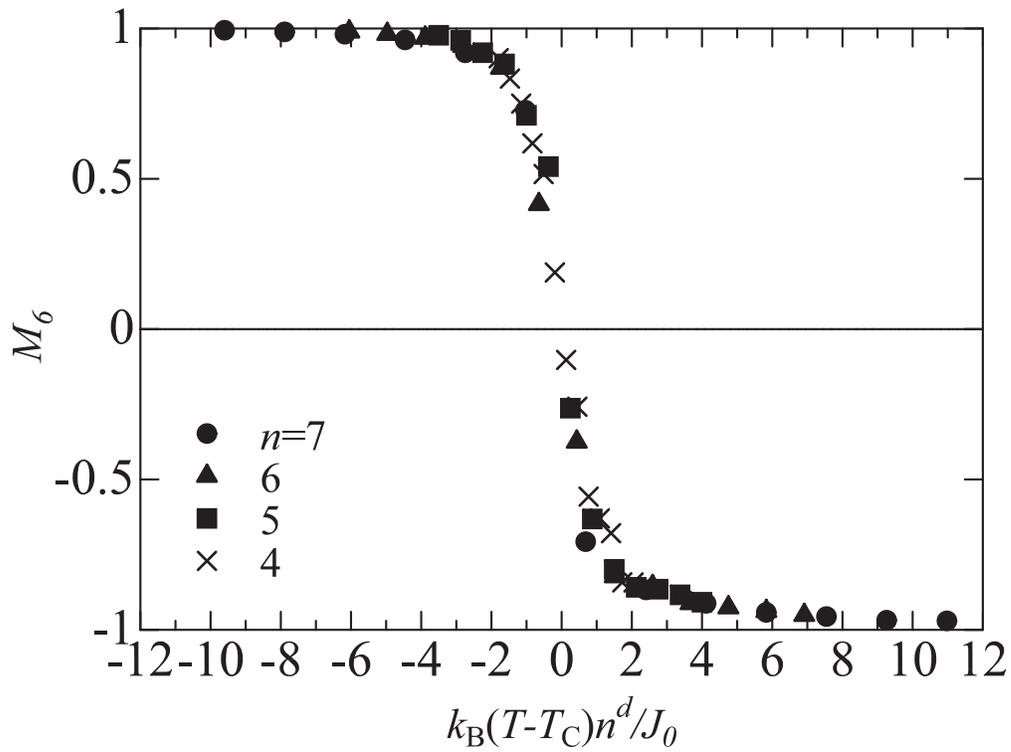}
\caption{Scaling plot of $M_6$ around the low temperature phase transition
 pont ($k_{\rm B}T_{\rm C}/J_0=0.188$) for the various system size.  \label{cos6theta}}
\end{figure}

We calculate the following quantity\cite{oshikawa,todoroki}:
\begin{eqnarray}
M_6=
\left \langle \cos \left (6\tan^{-1}\left ( \frac{m_y}{m_x}\right )\right )\right \rangle,
\end{eqnarray}
where
\begin{eqnarray}
m_x&=&m_1-\frac{1}{2}(m_2+m_3), \\
m_y&=&\frac{\sqrt{3}}{2}(m_3-m_2)
\end{eqnarray}
and $m_i$ ($i=1, 2, 3$) is the sublattice magnetization of $i$-th sublattice.
The present model can be transformed into a six-state clock model as depicted in Fig.\ref{clock}.
The quality 
\begin{eqnarray}
\theta =\tan^{-1}(m_y/m_x)
\end{eqnarray}
is the angle in the transformed model.
Then, $M_6=\langle \cos 6\theta\rangle$ detects the order of the six-state clock model
and it takes $1$ for the 2FR phase and $-1$ for the PD phase.
When the transition is of the first order, this quantity is scaled by the system size as
\begin{eqnarray}
\langle \cos 6\theta\rangle \sim f(k_{\rm B}(T-T_{\rm C})n^d/J_0),
\end{eqnarray}
around the transition point. 
Here $d$ is the dimension of the system, $T_{\rm C}$ is the transition temperature, and $f(x)$ is a 
finite size scaling function.
We obtained a good scaling plot with the $k_{\rm B}T_{\rm C}/J_0= 0.188$ as depicted Fig.\ref{cos6theta}. 
Therefore, we estimated that this transition is of the first order and the transition temperature is $k_{\rm B}T_{\rm C}/J_0=0.188(2)$.
In the high temperature region beyond $T_{\rm C}$, $M_6$ goes to $-1$, which means that the system is stable in the PD phase.
The macroscopic fluctuation which was proposed in the previous study\cite{koseki} does not appear.
The large fluctuation is attributed to large fluctuation due to the apparent 
XY behavior which is also observed in the 6GCL model. \cite{todoroki}
The 3FR phase is not observed. In the present study, we used a larger sized lattices.
Therefore we succeeded to detect 
bulk structures in the large fluctuation.
In Fig.\ref{cos6theta}, we see a good convergence of date for $n=6$ and $n=7$.

\begin{figure}
\includegraphics[scale=0.8]{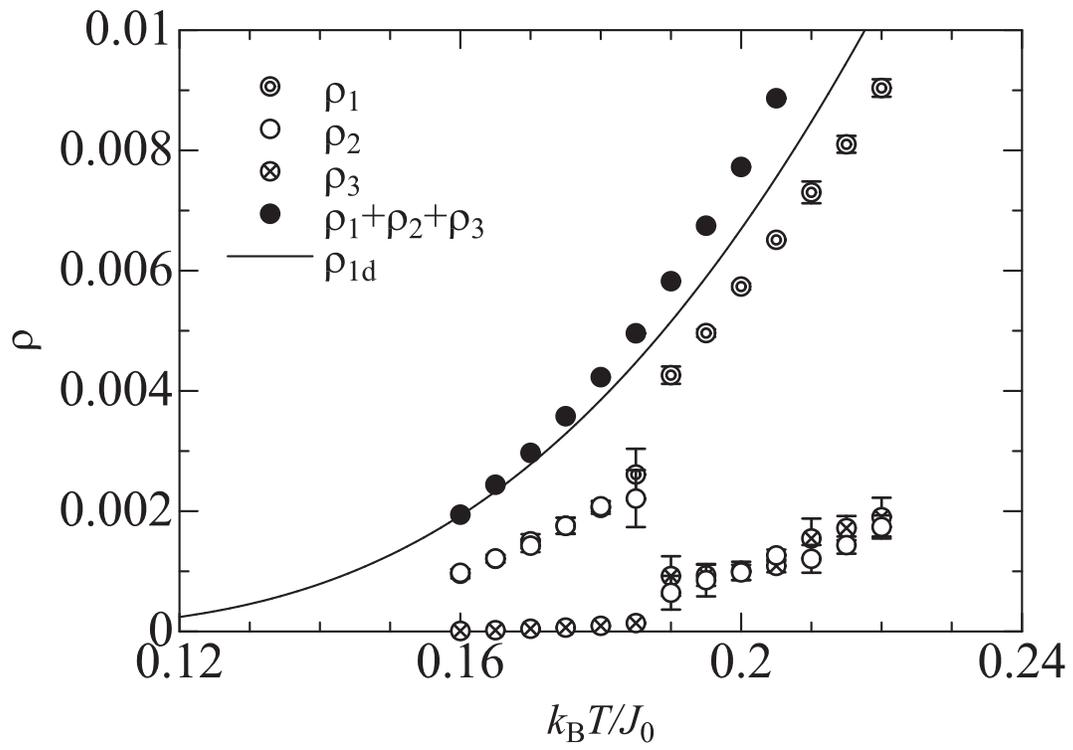}
\caption{Domain wall density along the c-axis in the each chain. 
$\rho_i$ is the domain wall dencity of $i$-th sublattice.\label{domainwall}}
\end{figure}

Although the low temperature phase transition is of the first order, 
the specific heat shows no anomaly at the transition point as depicted in Fig.\ref{domainwall}.
It means that the latent heat of this transition does not exist or at least is very small. 
The difference between the PD phase and the 2FR phase is whether the disordered chains exist, or not.
The energy difference between the ordered chain and the disordered chain is considered to be given mainly by 
the number of domain walls $n_{\rm d}$ in a chain. 
Because of the long correlation length along the c-axis, $n_{\rm d}$ is very small.
In order to study how the disordered chain changes to the ordered chain,
we examine the domain wall density
\begin{eqnarray}
\rho_i=n_{{\rm d},i}/L
\end{eqnarray}
in the $i$-th sublattice.
Here $L$ is the length of the chain. 
We show the temperature dependence of the density of domain walls for the case of $n=7$ in Fig.\ref{domainwall}.
In the PD phase, the inter-chain interactions cancel out each other on the disordered chain.
Therefore this chain can be regarded as a one dimensional Ising chain with interaction $J_0$,
when we ignore the next nearest neighbor interaction $J_2$ because 
$|J_2/J_0|=1\times 10^{-4}$.
The domain wall density of the disordered chains $\rho_1$
is roughly represented by that of the 
one-dimensional Ising model
\begin{eqnarray}
\rho_{\rm 1d}=\frac{e^{-\beta J_0/2}}{e^{\beta J_0/2}+e^{-\beta J_0/2}}.
\end{eqnarray}
In Fig.\ref{domainwall}, 
we find a discrete change around $k_{\rm B}T/J_0\simeq0.188$. The region $k_{\rm B}T/J_0\ge0.188$
corresponds to the PD phase. There, 
$\rho_{1}$ is a little smaller than $\rho_{\rm 1d}$ which is attributed to the next nearest neighbor interaction.
The other two chains are ordered chain and, 
the densities of the domain wall, $\rho_2$ and $\rho_3$ are small, 
and $\rho_{\rm tot}=\rho_1+\rho_2+\rho_3\simeq \rho_1$.
On the other hand, in the 2FR phase ($k_{\rm B}T/J_0< 0.188$), 
the domain walls move to the two less-ordered chains ($\rho_1$ $\simeq$ $\rho_2$). Thus, the configurations of the 
domain wall distribution changes discontinuously, which is consistent with the first order phase transition.
Here it should be noted that the $\rho_{\rm tot}$ changes very little at the transition point.
Thus, 
the latent heat (or the energy differences) between the PD phase and the 2FR phase
due to the domain wall is very small.
Therefore, the dominant contribution to the latent heat 
comes from energy due to the next nearest neighbor interactions.
For CsCoCl$_{3}$ and CsCoBr$_{3}$, 
the contribution due to the next nearest neighbor interaction $J_2$ is much smaller than that of the domain wall 
(in the present study $|J_2/J_0|=1\times 10^{-4}$).
Thus, it is difficult to measure of the anomaly of the specific heat 
in the experimental and numerical observations.

\section{Cluster Mean Field Approximation}
\begin{figure}
\includegraphics[scale=0.8]{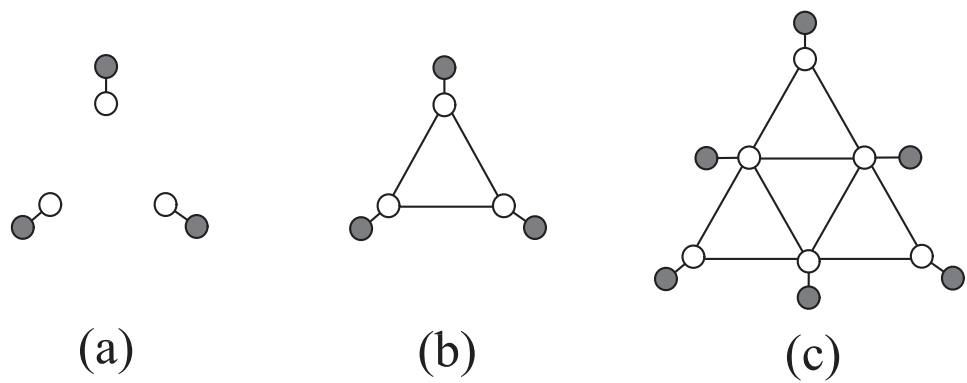}
\caption{Clusters for the mean field approximations with mean fields. The gray circles represent the mean fields and white circles represent the spins of the cluster.\label{cluster}}
\end{figure}

\begin{figure}
\includegraphics[scale=0.8]{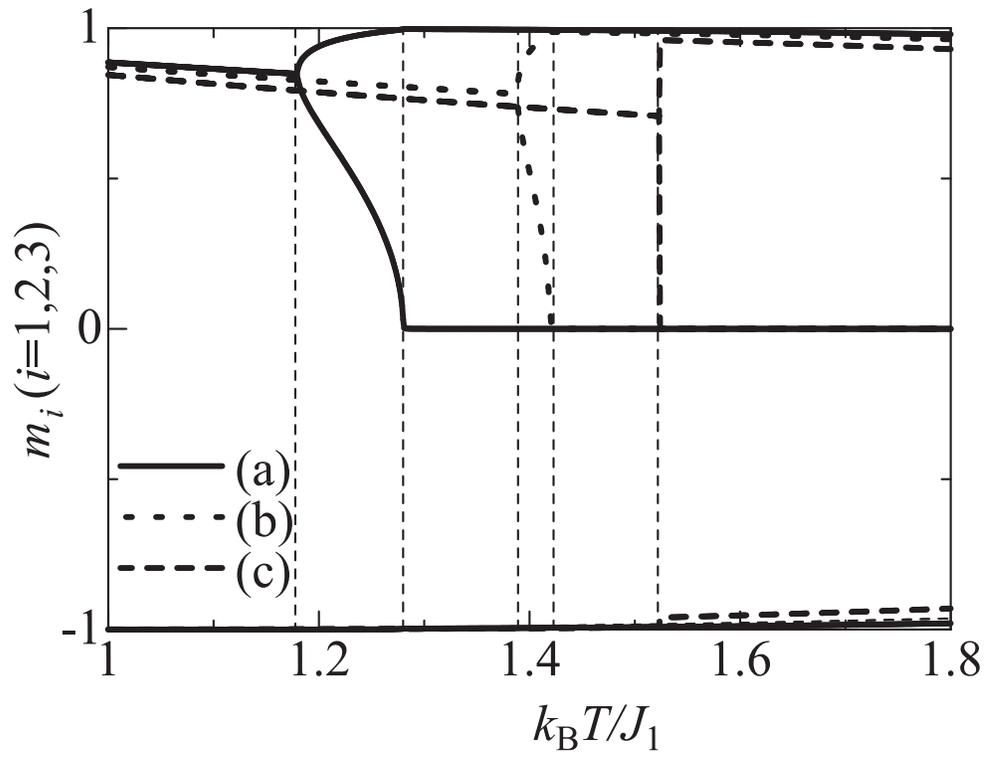}
\caption{Temperature dependence of the sublattice magnetizations obtained by the mean field approximations with clusteres depicted in Fig.\ref{cluster}. \label{mfa_submag}}
\end{figure}

As mentioned in the introduction, as to the existence of the 3FR phase, 
there is a discrepancy between the result of the MFA and of the Monte Carlo method.
In order to resolve this disagreement, we study the effect of fluctuation in the MFA.
In order to take the effect of fluctuation into account, we 
adopt so-called cluster mean field approximation where we investigate 
clusters of spin with the mean field representing
the interaction with the outer spins.
Therefore, fluctuations of spins inside the cluster are taking into account.
We takes the cluster depicted in Fig.\ref{cluster}.
The temperature dependence of the sublattice magnetizations are plotted in Fig.\ref{mfa_submag}.
In the model (a) we reproduced the Mekata's result. 
In the model (b) the region of the 3FR phase decreases, and in the model (c), 
it disappears. Therefore, we conclude that the 3FR phase does not stable against fluctuation.
The suppression of the 3FR phase has been reported in the studies using other extended mean field theory \cite{shiba, kurata}.
If the 3FR disappears, the PD phase can not change to the 2FR phase smoothly. 
Therefore we have the first order phase transition. These results are consistent with the results 
of MC simulation in the previous section.
\section{Conclusion and Discussion}
We study properties of ordered phases of the STAFI model.
At low temperature, the  present model has 2FR phase. Between the 2FR and the paramagnetic phase
at high temperatures, there is an intermediate phase of the spin structure of the PD phase.
There is no the 3FR phase predicted by a simple MFA.
This structure of phases is homomorphic to that of the 6GCL model, 
but not that of the isotropic six-state clock model.
In the PD phase, large fluctuation appears, which is similar to the IOP on the 6GCL model. 
This large fluctuation made confusion in previous studies.

The transition between the PD phase and the FR phase is of the first order.
However, no anomaly has been found in experiments and in simulations.
We have studied microscopic spin ordering in the chains along the c-axis to clarify the nature of the phase transition.
That is, we study whether the change at the low temperature critical point is
just a crossover from the PD type to the 2FR phase or a first order phase transition.
We found a discontinuous change of the distribution of domain walls
in the chains of the sublattices,
and found the change is of the first order.
However, the total number of domain wall changes very little at the point, which causes a very small singularity 
of the specific heat.

\begin{figure}
\includegraphics[scale=0.8]{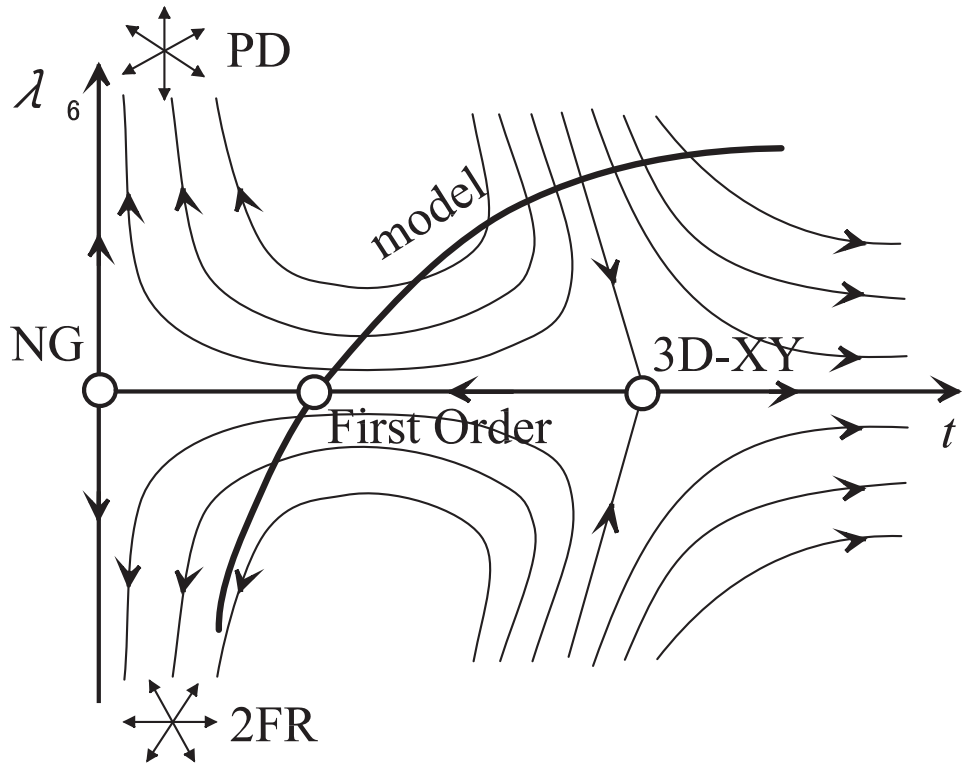}
\caption{Renormalization flow diagram of the Z$_6$ symmetry model which project the $t$-$\lambda_6$ surface. NG and 3D-XY means Nambu-Goldstone fixed point and 3D-XY fixed point, respectively.\label{renormalization}}
\end{figure}

From the viewpoint of the renormalization group, \cite{oshikawa,blankschtein}
the first order is also supported.
Suppose the effective Hamiltonian of the present model is
\begin{eqnarray}
{\cal H}=\int d^3{x}\left [
|\nabla \Phi|^2+t|\Phi|^2+u|\Phi|^4
+\sum_{n}\lambda_{6n}(\Phi^{6n}+\bar{\Phi}^{6n}) \right ],
\end{eqnarray}
where $\Phi$ is the complex field and 
$\bar{\Phi}$ is its conjecture. 
The renormalization group flow diagram of this model
is depicted in Fig.\ref{renormalization}. The 2FR phase and the PD phase correspond to $\lambda_{6} \rightarrow -\infty$
and $\lambda_{6}\rightarrow \infty$, respectively. 
Thus, the transition point between the PD and the 2FR phase is at $\lambda_6=0$.
The nature of this transition at $\lambda_6=0$ is similar to that of the XY model 
when the external field $\vec{H}$ change its sign.
Below the transition temperature, the first order phase transition exists at $\vec{H}=0$.
This first order transition of the XY model has a unique property that the 
correlation length diverges at $\vec{H}=0$.
However, in the present STIAF model, 
due to the higher terms of $\lambda_{6n}$($n\ge2$), 
it is expected that the transition becomes an ordinary first order transition where the correlation length is finite.
Therefore, the transition between the PD phase and the 2FR phase is of the ordinary first order.

We hope that the present study settles the long-studied problem of the low temperature transition of the STAFI model.

\section*{Acnowledgments}
The authors would like to express their sincere thanks to Professors Y. Ueno, F. Matsubara, M. Oshikawa
and Y. Ozeki for valuable discussions and comments. 
The numerical calculations were partially performed using the facilities
of the Supercomputer center, ISSP, University of Tokyo.

The present work is partially supported by Grant-in-Aid from the Ministry of Education, Culture,
Sports, Science and Technology.

\newpage

\newpage

\newpage

\newpage

\newpage

\newpage

\newpage

\newpage

\newpage

\newpage
\end{document}